\documentclass[prl,twocolumn,superscriptaddress]{revtex4}
\usepackage{graphicx}
\usepackage{amssymb}
\usepackage{amsmath}
\usepackage{xspace}

\newcommand{\ii}{\mathrm{i}}
\let \Re \relax
\DeclareMathOperator{\Re}{Re}

\begin{document}

\title{Boson-controlled quantum transport}

\author{A.~Alvermann}
\affiliation{
Institut f\"ur Physik, Ernst-Moritz-Arndt-Universit{\"a}t Greifswald, 17489 Greifswald, Germany }
\author{D. M. Edwards}
\affiliation{
Department of Mathematics, Imperial College London, London SW7 2AZ, United Kingdom
}
\author{H.~Fehske}
\affiliation{
Institut f\"ur Physik, Ernst-Moritz-Arndt-Universit{\"a}t
Greifswald, 17489 Greifswald, Germany }

\begin{abstract}
We study the interplay of collective dynamics
and damping 
in the presence of correlations and bosonic fluctuations
within the framework of a newly proposed model,
which captures the principal transport mechanisms that apply to a
variety of physical systems.
We establish close connections to the transport of lattice and
spin polarons, or the dynamics of a particle coupled to a bath.
We analyse the model 
by exactly calculating the optical conductivity, Drude weight,
 spectral functions, groundstate dispersion and particle-boson correlation
 functions for a 1D infinite system.
\end{abstract}

\maketitle

The motion of an electron or hole which interacts strongly with some
background medium is a constantly recurring theme in condensed matter
physics. Media which commonly occur are ordered spin
backgrounds as in the t-J model of doped Mott insulators
or vibrating lattices, as in the Holstein and 
quantized Su-Schrieffer-Heeger (SSH) models for 
polarons or charge density waves.
A moving particle creates local distortions 
of substantial energy in the medium, e.g. local spin
fluctuations, which may be able to relax. Their relaxation
rate determines how fast the particle can move.
In this sense particle motion is not free at all; the
particle is continuously creating a string of distortions but can move on
``freely'' at a speed which gives the distortions time to decay. 
This picture is very general with wide applicability, e.g.
to charge transport in high-$T_c$ superconducting and colossal
magnetoresistive materials, mesoscopic devices like quantum wires, and
presumably even biological systems.~\cite{ALL} 

In this Letter we investigate transport within some background medium
by means of an effective lattice model with a novel form of
electron-boson coupling.~\cite{Ed06}
The bosons correspond to local fluctuations of the background.
To motivate the form of the model let us consider a 
hole in a 2D antiferromagnet.
In a classical Neel background motion of the hole creates a string of
misaligned spins.
This `string effects' 
strongly restricts propagation.~\cite{BR70}
If however spins can exchange quantum-mechanically  
distortions of the spin background can `heal out' by 
local spin fluctuations with a rate controlled by the exchange
parameter $J$.
This way the hole can move coherently with a bandwidth
proportional to $J$.~\cite{KLR89}
Here we present a simpler spinless model which captures
this two-fold physics within a generic framework.

Specifically, we consider the Hamiltonian
\begin{equation}\label{Ham}
H =  - t_f \sum_{\langle i, j \rangle}  c_j^{\dagger} c^{}_i \;
-t_b \sum_{\langle i, j \rangle}  c_j^{\dagger} c^{}_i (b_i^{\dagger}+
b^{}_j) \;
+ \omega_0 \sum_i b_i^{\dagger} b^{}_i 
\end{equation}
for fermionic particles ($c^\dagger_i$) coupled to bosonic fluctuations
($b^\dagger_i$) of energy $\omega_0$. $H$ models two transport processes,
one of unrestricted hopping $\propto t_f$, 
and a second of boson-controlled hopping $\propto t_b$.
While for $t_b=0$ the model reduces to that of a free particle, for
$t_b \ne 0$ the physics of the model is governed by two ratios:
The relative strength $t_b / t_f$ of the two transport mechanisms,
and the rate of bosonic fluctuations $t_b / \omega_0$.
Therein the model also resembles common electron-phonon 
models like the Holstein or SSH model.
Nevertheless it differs in that particle hopping creates a boson 
only on the site the particle leaves, 
and destroys a boson only on the site the particle
enters. As a consequence the `string' effect familiar 
from the t-J model is captured within the model,
but also more complex features like in the 2D t-J model occur
already in a 1D setting.
In this contribution we study the model for a single particle in
1D at temperature $T=0$.

\begin{figure}
\includegraphics[width=0.8\linewidth]{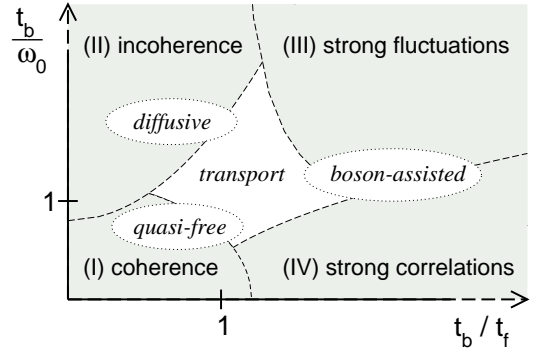}
\caption{Schematic view of the four physical regimes described by
  Hamiltonian~\eqref{Ham}, and the transport properties (see text).
}
\label{FigPhase}
\end{figure}

We begin with a discussion of the physical regimes shown in
Fig.~\ref{FigPhase}.
First, for small $t_b / t_f$ (left side),
transport takes place through unrestricted hopping.
There, the model essentially describes motion of a
particle coupled to a bosonic bath, when any bosonic fluctuations reduce the
mobility of the particle. 
For small $t_b/\omega_0$ (regime I),
the number of bosons is small. The particle propagates almost
coherently,  and transport resembles that of a
free particle.
If $t_b/\omega_0$ is larger (regime II), the number of bosons
increases, and the bosonic timescale is slower than
that of unrestricted hopping.
Therefore bosonic fluctuations mainly act as random, incoherent scatterers,
and the particle loses its coherence. The transport
is then diffusive, with a short mean free path.
In the second limiting case, for large $t_b / t_f$ (right side),
transport takes mainly place through boson-controlled hopping,
i.e. particle motion relies on the existence of bosons,
which are created and consumed in the hopping process.
For large $t_b / \omega_0$ (regime III),
transport is limited by strong scattering off uncorrelated 
bosonic fluctuations (similar to regime II).
For small $t_b / \omega_0$ however (regime IV), the bosons 
instantly follow the particle motion and strong correlations develop,
leading to collective particle-boson dynamics.
Note that boson-controlled hopping acts in two opposing ways:
Depending on how many correlations between the bosons persist,
it may either limit transport as a result of scattering
off random fluctuations (regimes  II and III), but may also enhance 
transport through correlated emission and absorption of bosons
(regime IV).

For large $t_b/t_f$ and small $t_b/\omega_0$, transport in the model
resembles that of the hole-doped t-J-model.~\cite{Tr88,KLR89}
To make this connection more explicit, we perform the unitary 
transformation $b_i \mapsto b_i - t_f / 2 t_b$ of
$H$:
\begin{equation}\label{HamTrans}
H' =  -t_b \sum_{\langle i, j \rangle}  c_j^{\dagger} c_i (b_i^{\dagger}
+ b^{}_j)  \\
 - \lambda \sum_i (b_i^{\dagger} + b_i) 
+ \omega_0 \sum_i b_i^{\dagger} b_i 
\end{equation}
(here a constant energy shift is dropped, which is
proportional to the number of lattice sites $N$ and guarantees finite results
for $N\to\infty$). In \eqref{HamTrans}  
the coherent hopping channel is eliminated 
in favor of a boson relaxation term 
$\lambda \sum_i (b_i^{\dagger} + b_i)$.
If $\lambda=\frac{\omega_0 t_f}{2 t_b}>0$,
i.~e. $t_f > 0$ in the original model,
the decay of bosonic excitations allows for 
$t$-$J$-like quasiparticle (hole) transport.
In the `classical' limit $\lambda\to 0$ coherent quasiparticle motion
is suppressed as in the $t$-$J_z$ (Ising spin) model.
Note that the limit $\omega_0 \to 0$ does not immediately lead to a
semi-classical description established for the Holstein- and SSH-model
since the electron does not couple exclusively
to oscillator coordinates $\propto (b^{}_i+b^\dagger_i)$.

\begin{figure}
\includegraphics[width=0.95\linewidth]{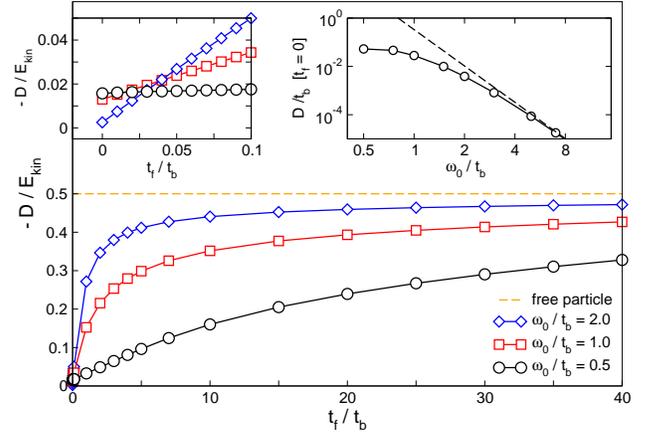} \\
\caption{(Color online) Drude weight $D$ scaled to the kinetic energy $E_\mathrm{kin}$.
The left inset displays the region $t_f \gtrsim 0$ magnified,
while the right inset shows $D$ for $t_f=0$ in dependence on
$\omega_0$. The dashed curve gives
 the asymptotic result
$D \simeq t_b^6 / (3 \omega_0^5) + O(t_b^8 / \omega_0^7)$
for $\omega_0 \to \infty$.
}
\label{FigDrude}
\end{figure}

For a quantitative analysis of transport
in the different regimes
we employ the optical conductivity
$ \Re \sigma(\omega) = 2 \pi D \,\delta(\omega) +
 \sigma_\mathrm{reg}(\omega)$,
where the regular part is 
\begin{equation}
  \sigma_\mathrm{reg}(\omega) = 
  \pi \sum_{n>0} \frac{|\langle n | j |0\rangle |^2}{\omega_n}
  \,[\delta(\omega - \omega_n) + \delta(\omega+\omega_n)] \;.
\end{equation}
Here $|n\rangle$ labels the eigenstates of the one-fermion system
with excitation energy $\omega_n = E_n-E_0$,
$|0\rangle$ is the groundstate.
The current operator $j = j_f + j_b$ is given by
\begin{align}
 j_f &=  \ii t_f \sum_{i} \,
  c^\dagger_{i+1} c^{}_i - c^\dagger_i c^{}_{i+1}\;, \\
 j_b &= \ii t_b \sum_{i} \,
  c^\dagger_{i+1} c^{}_i b^\dagger_i - c^\dagger_i c^{}_{i+1}
  b^{}_i  
- c^\dagger_{i-1} c^{}_i b^\dagger_i + c^\dagger_i c^{}_{i-1}
  b^{}_i   \;. \nonumber
\end{align}
The Drude weight $D$ serves as a measure of the coherent, free particle like
transport, and fulfils the f-sum rule
\begin{equation}\label{EqFSum}
  \int_{-\infty}^\infty \! \sigma(\omega) d\omega =
  2\pi D + 2 \int_0^\infty \! \sigma_\mathrm{reg}(\omega) d\omega =  
  - \pi \, E_\mathrm{kin} \;,
\end{equation}
where
$E_\mathrm{kin} = \langle 0|H-\omega_0 \sum_i b^\dagger_i
b^{}_i|0\rangle$ is the kinetic energy.
For a free particle ($t_b = 0$), the Drude weight
is given by $D = t_f$,
and the sum rule reads $- D / E_\mathrm{kin} = 0.5$,
while $- D / E_\mathrm{kin} \ll 0.5$ for diffusive transport in the
presence of strong fluctuations.
We can therefore characterize the different transport regimes through
the ratio $- D/E_\mathrm{kin}$ (Fig.~\ref{FigDrude}).
The curve for $\omega_0 / t_b =2.0$
shows that in a wide range of $t_f/t_b$
transport is quasi-free with $- D/E_\mathrm{kin} \lesssim 0.5$.
For smaller $\omega_0 / t_b$, as the number of fluctuations is larger,
$-D/E_\mathrm{kin}$ is decreased due to scattering.
The smaller $\omega_0 / t_b$, the slower
$-D/E_\mathrm{kin}$ tends to its asymptotic values $0.5$ for
$t_f/t_b\to\infty$.
This shows
how the crossover from the coherent regime (I) with quasi-free transport
to the incoherent regime (II) with diffusive transport  is controlled by
$t_b/\omega_0$.
For small $t_f/t_b$, when boson-controlled hopping is the dominating
transport process, $D$ increases with decreasing
$\omega_0$ (left inset).
This is the regime of boson-assisted transport, where transport is
mediated by vacuum-restoring processes (see below).
For the moment, we note, that $D$ at $t_f=0$ saturates with $\omega_0 \to
0$ (right inset), as one passes from the correlation dominated
regime (IV) to the regime (III) of strong, uncorrelated fluctuations.

We complete our study by means of three quantities:
First, the groundstate dispersion $E(k)$ provides 
the effective mass 
$1/ m^* = \frac{\partial^2 E}{\partial k^2}|_{k=0}$,
which is renormalized for $t_b \ne 0$.
By Kohn's formula, $D = 1/(2 m^*)$.
Second,
we use the particle-boson correlation function
\begin{equation}
  \chi_{ij} = \langle 0| b^\dagger_i b^{}_i c^\dagger_j c^{}_j  | 0 \rangle \;,
\end{equation}
and third, the one-particle
spectral function
\begin{equation}
\label{aspekt}
A(k,\omega) 
=  \sum_n 
|\langle n|c^\dagger_k 
|\mathrm{vac}\rangle|^2 \,\delta (\omega-\omega_n)  \;,
\end{equation}
where $|\mathrm{vac}\rangle$ denotes the particle vacuum.
All quantities have been calculated 
with exact numerical techniques (Lanczos diagonalization and kernel
polynomial methods~\cite{WWAF06}).
We employed a basis construction for the many-particle Hilbert
space that is variational for an infinite lattice.~\cite{BTB99}
This guarantees data of extremely high precision,
free of finite size effects.

\begin{figure}
\includegraphics[width=0.95\linewidth]{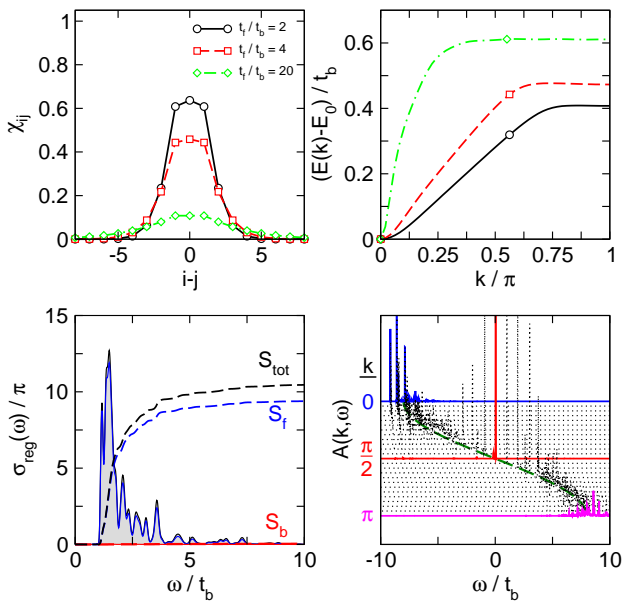}
\caption{(Color online)
Top row: $\chi_{ij}$ (left)
and $E(k)-E_0$ (right)
for  $\omega_0 / t_b = 0.5$ and $t_f / t_b = 2,4,20$.
Bottom row: $\sigma_\mathrm{reg}(\omega)$ (left) for $t_f / t_b = 20$
and $A(k,\omega)$ (right) for $t_f / t_b = 4$.
To analyze the relative importance of the two transport processes 
$j_f$ and $j_b$,
we show the corresponding contributions
$\sigma_f(\omega)$, $\sigma_b(\omega)$ to
$\sigma_\mathrm{reg}(\omega)$ separately
(note that generally $\sigma_\mathrm{reg}(\omega) \ne
\sigma_f(\omega)+\sigma_b(\omega)$).
$S_\mathrm{tot}$ (and similar $S_f$, $S_b$) denotes the integrated
conductivity $S_\mathrm{tot}(\omega)=\int_0^\omega \sigma_\mathrm{reg}(\omega') d\omega'$.
}
\label{FigFluct}
\end{figure}

Let us first discuss the `incoherent' or `diffusive' regime (II) (see
Fig.~\ref{FigFluct}).
As expected, the regular part of the optical conductivity is dominated
by a broad incoherent absorption continuum,
and $- D/E_\mathrm{kin}$ is small.
This resembles the situation for a large Holstein (lattice) polaron,
where the role of bosonic fluctuations is taken by optical phonons.~\cite{BWF98}
$\chi_{ij}$ shows that bosonic fluctuations are rather weakly correlated.
Of course, they form a cloud surrounding the particle, 
but are not further correlated.
The spectral function $A(k,\omega)$ supports this picture.
The spectral weight is distributed along the `free' dispersion
$-2 t_f \cos k$, like for a weakly bound particle-boson excitation.
Around $k=0$ and $k=\pm\pi$, the overdamped character of 
particle motion is very prominent.
Comparing with $E(k)$, we see that the
quasi-particle weight is negligible away from $k=0$,
and a well-defined quasi-particle band does not exist.
Somewhat surprisingly, for $k=\pm\pi/2$ almost all weight resides in a single
coherent peak at $\omega=0$.
A particle injected with $k=\pm\pi/2$ therefore propagates almost
unaffected by bosonic fluctuations.
In a sense, the system is transparent at this energy,
similar to e.g. a double well at certain resonant energies.

\begin{figure}
\includegraphics[width=0.95\linewidth]{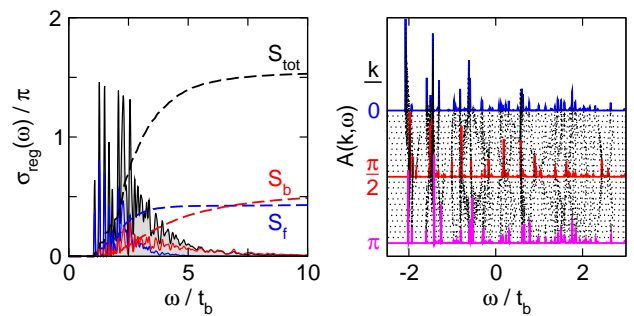} 
\caption{(Color online)
$\sigma_\mathrm{reg}(\omega)$ 
for $\omega_0 / t_b = 1.0$, $t_f / t_b = 2.0$ (left),
and $A(k,\omega)$ for 
$\omega_0 / t_b = 0.5$, $t_f / t_b = 0.04$ (right).}
\label{FigInter}
\end{figure}

\begin{figure}
\includegraphics[width=0.95\linewidth]{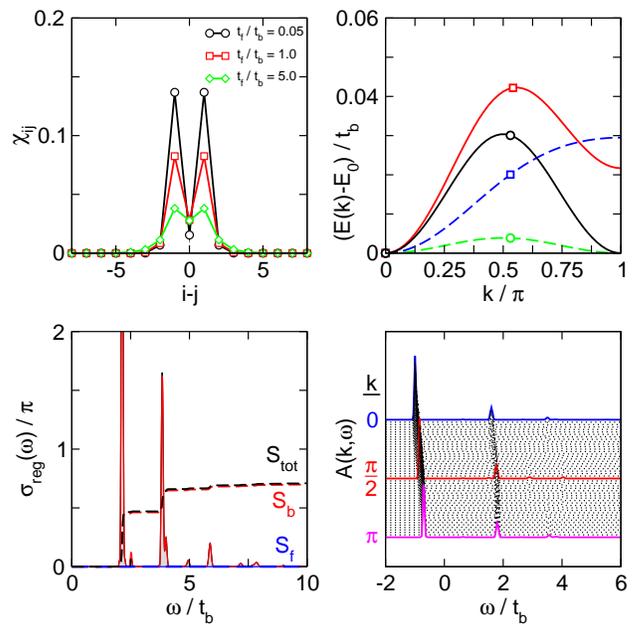}
\caption{(Color online)
Top left: $\chi_{ij}$ 
for  $\omega_0 / t_b = 2$ and $t_f / t_b = 0.05$ (circles), $1.0$
(squares), $5.0$ (diamond).
Top right: 
$E(k)-E_0$ 
for $\omega_0 / t_b =1$ (solid) and $\omega_0 / t_b =2$ (dashed),
with $t_f / t_b = 0$ (circles), $t_f / t_b =0.01$ (squares).
Bottom row: $\sigma_\mathrm{reg}(\omega)$ (left)
and $A(k,\omega)$ (right) for $\omega_0 / t_b = 2$,
$t_f / t_b = 0.1$.}
\label{FigCorr}
\end{figure}

\begin{figure}
\includegraphics[width=\linewidth]{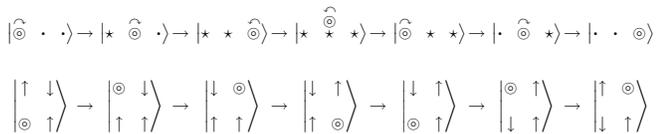}
\caption{Lowest order vacuum-restoring process (upper row),
which is a one-dimensional representation of the `Trugman path'~\cite{Tr88} in a N\'{e}el-ordered
spin background (lower row).
In steps 1--3, three bosons are excited, which are consumed
in steps 4--6.
Note how the correlated many-particle vacuum of the spin model is
translated to the bosonic vacuum.
}
\label{FigEchter}
\end{figure}

From the diffusive regime, we can evolve in two directions.
First, if we increase $t_b/t_f$ ($\leadsto$ regime (III)),
the contribution of boson-controlled hopping to the conductivity 
begins to dominate (see Fig.~\ref{FigInter}, left panel).
Still, transport is mainly diffusive (cf. Fig.~\ref{FigDrude}).
If we further increase  $t_b/t_f$ while keeping $t_b/\omega_0$ large
 (inside regime (III)),
strong but uncorrelated bosonic fluctuations develop. 
As a result, the spectral function $A(k,\omega)$ becomes fully
incoherent (right panel),
and $- D / E_\mathrm{kin}$ is small.
Apparently, the large number of bosonic fluctuations prevents
strong correlations (cf. $\chi_{ij}$ in Fig.~\ref{FigFluct}).
In the second direction, for large $t_b/t_f$ and small $t_b/\omega_0$
(regime (IV)), the number of fluctuations is reduced.
Then, 
strong correlations evolve
(see $\chi_{ij}$ in Fig.~\ref{FigCorr}).
The conductivity is now entirely given by the contribution from boson-controlled
hopping, but does not show the absorption continuum we found for
diffusive transport
(note that, although $ - D / E_\mathrm{kin}$ must be small for large $t_b
/ t_f$, it is much larger than for larger $t_b / \omega_0$).
Instead, both $\sigma_\mathrm{reg}(\omega)$ and $A(k,\omega)$ consist
of a few, well separated peaks.
This indicates, that here the model shows collective particle-boson dynamics,
i.~e. a well-defined quasi-particle exists,
like a spin/magnetic polaron in the t-J-model.~\cite{KLR89,SWZ88}
As a particular feature of the correlated transport mechanism which
dominates for  $t_b / t_f \gg 1$,
the quasi-particle dispersion $E(k)$ develops a $k\to k+\pi$ symmetry
for $t_b / t_f \to \infty$.
At $t_f=0$ the model therefore shows an electronic topological
transition, for which the hole doped t-J-model provides a
specific example.

The correlated transport mechanism for  $t_b / t_f \gg 1$ is best
understood in the limit $t_f=0$.
Then, the particle can only move by creating bosonic
fluctuations, i.e. transport is fully boson-assisted.
Moving by $L$ sites costs energy $L\omega_0$, 
which is just the string effect and tends to localize the particle.
However, there exist processes that propagate the particle but restore
the boson vacuum in their course.
The lowest order process of this kind  comprises 6
steps, promoting the particle by 2 sites (Fig.~\ref{FigEchter}).
By this and similar higher order processes the particle is itinerant
even at $t_f=0$, with a finite, though small, Drude weight.
Since in any hop the boson number changes by one, any vacuum-restoring
process propagates the particle by an even number of
sites.
This immediately explains,
why $E(k)$ for $t_f=0$ has period $\pi$.
The weight of the lowest order process shown in Fig.~\ref{FigEchter}
scales as $t_b^6 / \omega_0^5$ (cf. Fig.~\ref{FigDrude}).
Accordingly, boson-assisted transport dominates for large
$ (t_b / \omega_0)^5 (t_b / t_f)$.
In this regime, the mobility of the particle increases if $\omega_0$
decreases, as vacuum-restoring processes become energetically more favorable.
This explains the opposite dependence of $D$ on $\omega_0$ for
$t_f/t_b \ll 1$ and $t_f/t_b \gtrsim 1$.
In the plot of the Drude weight
the transition from uncorrelated, diffusive to
correlated, boson-assisted transport therefore leads to a crossing of
the curves at about $t_f/t_b \sim 0.05$ (see Fig.~\ref{FigDrude}, left inset).
This substantiates our initial remark
that bosonic fluctuations act in two competing
ways: While they limit transport by unrestricted hopping if strong
but weakly correlated, they assist transport by boson-controlled
hopping in the regime of strong correlations.
Equally important, the same boson-assisted mechanism which allows for
transport is limited by itself,
because strong correlations cannot persist in the presence of many
fluctuations:
The Drude weight $D_{t_f=0}$ for $t_f=0$ 
increases with decreasing
$\omega_0$ as we discussed, but finally saturates.
The large number of bosonic fluctuations, which
become increasingly uncorrelated for small $\omega_0$, interfere with
the correlated boson excitations in the vacuum-restoring transport processes,
restricting their efficiency.
Note that these fundamental physical mechanisms are not enforced in the
model, but emerge naturally from the competition between uncorrelated
scattering and correlation-induced transport.

To conclude,
the model that we propose provides
a reduced but realistic description of fundamental aspects of
transport in the presence of bosonic fluctuations.
Topics of current interest
become accessible within the comprehensive framework established.
Despite its seemingly simple form,
the model covers very different physical regimes, 
identified e.g. by the Drude weight.
For these regimes,
we establish links to
important many-body systems like lattice or spin polarons,
and could thereby demonstrate the relevance of the model.
Based on these results,
the model will certainly prove useful for further studies,
e.g. on the possibility of pairing for two particles, 
or on topological phase transitions for finite densities.
The whole phase diagram for finite density and temperature is open for
investigation.

We acknowledge helpful discussions with G.~Wellein and financial
support by DFG through SFB 652.

\end{document}